\documentclass[conference]{IEEEtran}
\IEEEoverridecommandlockouts
\usepackage{cite}
\usepackage{amsmath,amssymb,amsfonts}
\usepackage{graphicx}
\usepackage{subfig}
\usepackage{textcomp}
\usepackage{xcolor}
\usepackage{algorithm}
\usepackage{algpseudocode}
\usepackage{multirow}
\usepackage{caption}
\def\BibTeX{{\rm B\kern-.05em{\sc i\kern-.025em b}\kern-.08em
    T\kern-.1667em\lower.7ex\hbox{E}\kern-.125emX}}

\makeatletter
\newcommand{\linebreakand}{%
  \end{@IEEEauthorhalign}
  \hfill\mbox{}\par
  \mbox{}\hfill\begin{@IEEEauthorhalign}
}
\makeatother
\begin{document}

\title{\textit{CloudScent}: a model for code smell analysis in open-source cloud}
\author{\IEEEauthorblockN{ Raj Narendra Shah}
\IEEEauthorblockA{\textit{Department of Computer Science} \\
\textit{California State University}\\
San Marcos, California, USA \\
shah068@csusm.edu}
\and
\IEEEauthorblockN{ Sameer Ahmed Mohamed}
\IEEEauthorblockA{\textit{Department of Computer Science} \\
\textit{California State University}\\
San Marcos, California, USA \\
moham135@csusm.edu}
\and
\IEEEauthorblockN{ Asif Imran}
\IEEEauthorblockA{\textit{Department of Computer Science} \\
\textit{California State University}\\
San Marcos, California, USA \\
aimran@csusm.edu}
\linebreakand
\IEEEauthorblockN{Tevfik Kosar}
\IEEEauthorblockA{\textit{Department of Computer Science and Engineering} \\
\textit{University at Buffalo}\\
Buffalo, New York, USA \\
tkosar@buffalo.edu}
}
\maketitle
\begin{abstract}
The low cost and rapid provisioning capabilities have made open-source cloud a desirable platform to launch industrial applications. However, as open-source cloud moves towards maturity, it still suffers from quality issues like code smells. Although, a great emphasis has been provided on the economic benefits of deploying open-source cloud, low importance has been provided to improve the quality of the source code of the cloud itself to ensure its maintainability in the industrial scenario.   Code refactoring has been associated with improving the maintenance and understanding of software code by removing code smells. However, analyzing what smells are more prevalent in cloud environment and designing a tool to define and detect those smells require further attention. In this paper, we propose a model called \textit{CloudScent} which is an open source mechanism to detect smells in open-source cloud. We test our experiments in a real-life cloud environment using \textit{OpenStack}. Results show that \textit{CloudScent} is capable of accurately detecting 8 code smells in cloud. This will permit cloud service providers with advanced knowledge about the smells prevalent in open-source cloud platform, thus allowing for timely code refactoring and improving code quality of the cloud platforms. 
\end{abstract}


\section{Introduction}
Cloud computing is the dynamic provisioning of resources from a shared resource pool, assigned as per the demand of its users \cite{armbrust2010view}. The dynamic nature and rapid resource provisioning capability of the cloud have made it a desirable platform for a large group of applications. In addition to commercial cloud services, open-source cloud computing platforms are reaching maturity, and organizations are increasingly incorporating open-source cloud into their IT infrastructure. Organizations want to harness the power of economic benefits of open-source cloud to launch their complex applications. As a result, they require the cloud infrastructure to be maintainable and reliable.

The maintainability and acceptability of an open-source cloud platform are challenged if code smells are present in it \cite{antal2018hands}. Code smells are behavior in source code that indicate a deeper problem \cite{10.1145/3571697.3571704}. Refactoring code smells is the process of removing smells by restructuring code while preserving its functionality. Recent studies have focused on detecting code smells by parsing through user comments in cloud user groups \cite{tahir2020large}. The negative impact on resource consumption due to the presence of code smells has been studied. Semantic approaches to code smell detection in the cloud deployment code have been explored \cite{han2021understanding} However, a tool that detects code smells specific to open-source cloud needs to be identified.

In this paper, we provide a rule-based syntactic approach called \textit{CloudScent} to detect code smells in the critical modules of the open-source cloud. We use industry-standard rules to describe the smells and design an open-source tool to detect the smells. We explore the 4 key modules of the \textit{OpenStack} cloud namely \textit{nova, keystone, horizon, and neutron}. The smells are \textit{repetitive code, dead code, multiple returns, long statements, same function names, long classes and methods, long parameter list}, and \textit{long conditionals and loops}. We test our model using the real-life case study mentioned above. 

The rest of the paper proceeds as follows: Section \ref{related} discusses the related work in this area. Section \ref{methodology} explains the methodology of the research. Section \ref{model} presents the model of code smell detection in cloud. Section \ref{experiments} highlights the analysis and findings of the tool. Section \ref{threats} identifies the threats to validity of this research, and Section \ref{conclusion} concludes the paper.

\section{Related Work}
\label{related}
Current research on software code smells focuses on using tools like \textit{JDeodorant} \cite{tsantalis2018ten}, \textit{PMD} \cite{pmdsourcecodeanalyzer2022}, and \textit{SonarQube} \cite{marcilio2019static} that implement static analysis of code to detect and refactor smells. Although refactoring smells is important, what code smells need to be refactored with greater priority is dependent on the environment of the software which needs to be taken under consideration. Most of the existing tools are tested using offline software code which is executed in a single-user environment. However, how these tools perform in the case of distributed applications like the cloud need to be investigated.

\textit{Abstract Syntax Tree (AST)} based code smell detectors have been proposed to detect smells in Java applications \cite{moha2009decor}. A similar tool was proposed in \cite{10.1145/3571697.3571704}, and both papers focused on feedback from software engineers to identify the correctness of the detection operation. However, both papers used only two in-house applications to analyze the performance of the proposed tool. The performance of such tools in terms of preserving the correctness of code during refactoring was not studied. Also, the tool was not tested for real-life distributed applications such as the cloud.

\begin{table*}
\centering
\begin{tabular}{ p{5cm} p{9cm} } 
 \hline
 \textbf{Smell} & \textbf{Definition} \\ \hline 
Repetitive Codes & Group of a constant number of lines repeating in code. \\
 \hline 
 Dead Codes & The same block (conditional/loop) having another statement with a return. \\
 \hline 
 Multiple Return Statements & Having more than one return statement within a function.\\
 \hline 
 Long statements & A line having more than a constant number of words and characters.\\
 \hline 
 Multiple Same Function Names & Same name functions with the same number and types of parameters.\\
 \hline
 Long Classes Or Methods & A Class/function having more than a constant number of lines.\\
 \hline
 Long Conditionals or Loops & A conditional block having more than a constant number of lines.\\
 \hline
 Long Parameter list & A function having more than a constant count of parameters.\\
 \hline
\end{tabular}
\label{Codesmells3}
\caption{Identification of 8 code smells detected in OpenStack cloud environment}
\end{table*}

A recent investigation was conducted to understand which code smells were detected during code reviews and what actions reviewers suggested and developers took in response to the identified smells \cite{han2021understanding}. The authors obtained code reviews of two key modules of \textit{OpenStack} namely \textit{nova} and \textit{neutron}. The code reviews were used to detect code smells. The researchers determined that reviewers provided constructive feedback, including refactoring recommendations that helped developers remove smells. Although the research was conducted using \textit{OpenStack}, it depended mostly on user reviews who might not know all the different types of smells in the cloud. Hence, an expert tool was required which could cater to the requirements of code smells in the cloud.

The complexity of deploying cloud applications can cause cloud developers to inadvertently violate standard coding practices and introduce code smells \cite{kumara2020towards}.  The authors proposed a semantic approach that enables developers to address code smells that violate the modular structure of code. They used \textit{SPARQL}-based rules to identify the smells and tested their solution on a cloud prototype. Although this research is important, it should be tested on real-life cloud infrastructure like \textit{OpenStack}.

Current research had focused on detecting and annotating code smells for different types of systems which included programming languages, android, artificial intelligence platforms, data-intensive systems, and web applications. It had been stated that the flexible and multi-purpose nature of \textit{JavaScript} programming language introduced specifically 14 types of code smells \cite{9118465}. \textit{SQL-based} code smells were dominant in data-intensive systems due to their nature of frequently accessing database systems \cite{muse2020prevalence}. In another study, authors considered code smells that were present in 3 stack-exchange sites \cite{tahir2020large}. Focus had also been provided to analyze code smells in web-based e-commerce applications \cite{bessghaier2020diffusion}. Although existing research had provided an emphasis on various application platforms, significant efforts needed to be provided toward detecting and analyzing code smells which are prevalent in cloud computing.

\section{Methodology}
\label{methodology}

The goal of this paper is to detect which code smells occur in an open-source cloud. In this regard, we propose \textit{CloudScent}, which uses pre-specified rules to identify smells in open-source cloud platforms. To determine the code smells in cloud environment using \textit{CloudScent}, we selected the \textit{OpenStack} cloud, which was an open-source cloud computing platform \cite{rosado2014overview}. We surveyed the existing code smells and selected 8 such code smells that impacted cloud environments. We aimed to detect the cloud smells using the detector and determined whether the tool successfully detected those smells. Our research questions were as follows:

\begin{enumerate}
    \item Which code smells are prevalent in open-source cloud computing environments?
    \item What is an effective mechanism to detect the smells present in the source code of real-life cloud?
\end{enumerate} 

Our goal was to identify how code smells could be detected in a cloud environment through and effective model called \textit{CloudScent}. We tried to find out if any smell was more prevalent than the others. We presented our proposed approach in the next section.
\begin{figure*}[ht]
    \centering
    \includegraphics[height = 9cm, width=1\textwidth]{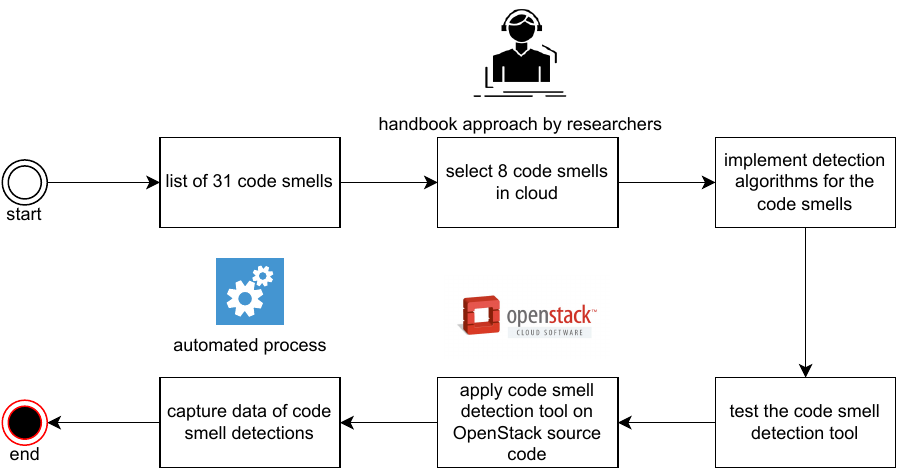}
    \caption{Flow of code smell detection mechanism for CloudScent.}
    \label{fig:mesh1}
\end{figure*}

\section{Proposed model}
\label{model}

The 8 code smells considered in the research have been defined in Table I. We explored the 31 code smells identified by software engineers \footnote{https://pragmaticways.com/31-code-smells-you-must-know/}. From there, we filtered 8 code smells that we considered to be significant for the cloud environment. This selection was made based on the syntax of the cloud in terms of multi-threading, dynamic resource provisioning, and elasticity. We designed a model for cloud computing code smell detection for \textit{OpenStack}. 

In this paper, we described the Algortihms of 3 different code smell detection processes of \textit{CloudScent} and the full source code of the detector was provided in \textit{Github}\footnote{https://github.com/rajs80266/CodeSmellDetection}. For all the 8 code smells, mock code lines were added in a file for parallel checking of accurate detection of individual smells. Based on the accuracy of the model, it was applied in the source code of \textit{OpenStack}. 

Figure \ref{fig:mesh1} showed the flow of the code smell detection mechanism applied to the \textit{OpenStack} code. We implemented the algorithms for the code smell detector. Then we tested the code smell detector on sample test files. Afterward, we applied the detector on the source code of \textit{OpenStack} cloud. We then stored the results of the detected code smells and represented those diagrammatically.

\begin{algorithm}
\caption{Obtain details of code units to be analyzed in CloudScent}\label{alg:cap}
\begin{algorithmic}
\State $DEFINE\ getFunctionName(line)$
\State $ SET\ fName = line.split()[1] $
\While{$detector \subset function $}
\If{$smell$ := function}
    \State $SET\ fName ]\in  fName[:fName.index('(')]$
\EndIf
\EndWhile \\
\Return $ fName$
\end{algorithmic}
\end{algorithm}
We considered a function as the smallest unit of code. In Algorithm \ref{alg:cap} we defined the function which we used to gather information regarding the units of code which contained the smells. We used this mechanism and detected the functions in the python files of the \textit{OpenStack} project which contained smells. Using the \textit{getFunctionName()}, we gathered important information regarding the function under scrutiny such as un-commented lines of code (\textit{loc}), the number of variables in the function, the number of parameters, etc. If the detector identified a function by matching its syntax to that of a smell, then it stored the function name and location, as well as recorded what smell type was present.
\begin{algorithm}
\caption{Checking the presence of Dead Code smell in CloudScent}\label{alg:deadcode}
\begin{algorithmic}
\State $SET\ nOfLines \leftarrow len(code)$
\State $SET\ leadSpaces \leftarrow getLeadSpaces(code[i])$
\State $SET\ j \leftarrow i + 1$
\While{$j \lessapprox nOfLines $}
\While {$leadSpaces = getLeadSpaces(code[j]))$}
\State $j += 1$
\If{$i+1 \neq j$}\\
    \Return $descSmell('Dead\ Code\ in', i + 1, j - 1)$
\EndIf
\EndWhile
\EndWhile
\end{algorithmic}
\end{algorithm}

Here, we also showed the pseudocodes for code smells namely \textit{dead code, long statements}, and \textit{long loops} and \textit{long conditionals}. In total, we set rules for detecting 8 types of code smells. For dead code smell, we initially collected the \textit{loc} inside each function. The details are shown in Algorithm \ref{alg:deadcode}. Next, the units of code are indexed and stored. We searched for lean spaces in the code, and once we got it, we recorded the lead space, and we compared the values to find if the recorded lead spaces were equal to the one-increment, if this is not the case, we reported a dead code smell.

\begin{algorithm}
\caption{Checking the presence of Long Statement smell using CloudScent}\label{alg:longstatement}
\begin{algorithmic}
\State $DEFINE\ chkLongStatements(code, i)$
\If{$len(code[i].lstrip().split()) > 20$}\\
    \Return $descCodeSmell('Long\ statement\ found', i, i)$
\EndIf
\end{algorithmic}
\end{algorithm}

To detect long statements, we started by removing the leading white spaces. Next, we recorded the individual statements using the \textit{split()} function. Each statement is considered an element and detected using the \textit{code[i].lstrip().split()}. The process was followed by counting the number of split statements. As a result, long statements were successfully detected.

\begin{algorithm}
\caption{Checking the presence of Long Blocks in CloudScent}\label{alg:longblocks}
\begin{algorithmic}
\State $DEFINE\ chkLongBlocks(code, i, blockType)$
\State $SET numberOfLines \leftarrow len(code)$
\State $SET\ leadSpaces \leftarrow getLeadSpaces(code[i])$
\State $SET\ j \leftarrow i$
\If{$blkType := "CLASS" \land (j - i) > 60$}\\
    \Return $descSmell('Long\ Class\ found', i, j - 1)$
\ElsIf{$blkType := "METHOD" \land (j - i) > 40$}
    \Return $descSmell('Long\ Method\ found', i, j - 1)$
\ElsIf{$blkType := "LOOP" \land (j - i) > 20$}
    \Return $descSmell('Long\ Loop\ found', i, j - 1)$
\ElsIf{$blkType := "CONDITIONAL" \land (j - i) > 10$}
    \Return $descSmell('Long\ Conditional\ found', i, j - 1)$
\EndIf
\end{algorithmic}
\end{algorithm}

In Algorithm \ref{alg:longblocks}, we presented the mechanism to detect code smells like long loops and long conditionals which we found to be present in significant numbers in the cloud. We considered the python files and we obtained each code block using the mechanism discussed in Algorithm 1. To check for the long loops, we took a block of code and matched it with the syntax of a loop in Python. Once a loop was determined, we checked whether the number of lines in the loop was greater than 20 or not. If the \textit{loc} in the loop is greater than a pre-specified number, we called it a long loop code smell.  A similar mechanism was considered for long conditional statements, with the exception that the conditional statements need to be longer than 10 uncommented \textit{loc} to be considered as a smell.

\section{Experiments and results}
\label{experiments}
The cloud environment had been specified to be \textit{OpenStack}. It was an open-source cloud platform that could be used by the cloud service provider to launch different cloud services like \textit{PaaS, SaaS, and IaaS}. For this research, we used the \textit{OpenStack Zed} series. We obtained the python code files for the 4 main components of \textit{OpenStack Zed} namely \textit{nova, neutron, horizon, and keystone} \cite{wen2012comparison}. The latest repository was obtained on \textit{19 February 2023} for this research. 
\begin{figure}[ht]
    \centering
    \fbox{\includegraphics[height = 6cm, width=0.44\textwidth]{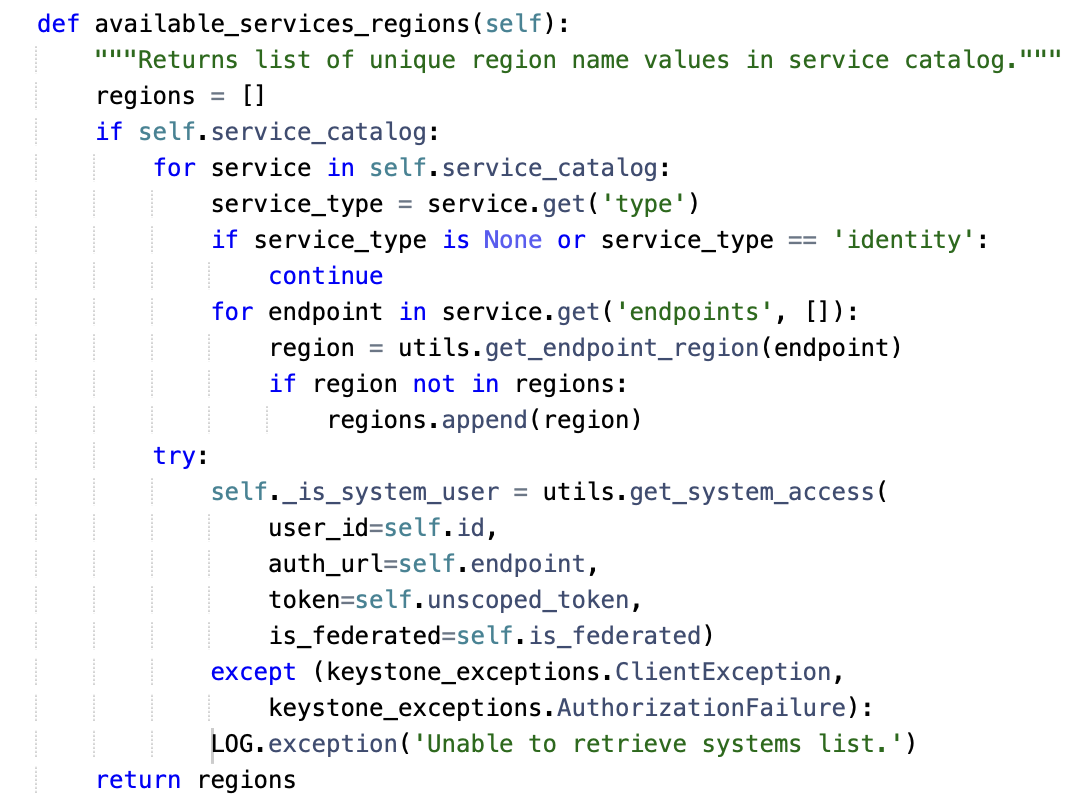}}
    \caption{Long conditional code smell obtained from OpenStack keystone project detected by CloudScent.}
    \label{fig:longconditional}
\end{figure}
Since cloud computing had become a significant paradigm where many real-life applications were executed, we selected the open-source cloud computing platform for our study. Cloud also offered a ubiquitous and parallel infrastructure to run critical scientific applications which were resource-intensive \cite{apostu2013study}. Detecting code smells in open-source cloud platforms like \textit{OpenStack} was an important activity since it provided insight to the cloud service providers on which smells to consider for the cloud.

While Algorithm 1 was parsing the \textit{OpenStack} cloud code files, some of the files had noises in them. These noises were moreover presented as comments in form of emoticons or similar typographs used by the \textit{OpenStack} programmers that caused our parsing script to fail. These noises were removed from files manually. The first author mainly took the lead in the removal of these noises. Any doubt during the removal was discussed between the first and second authors and any doubts were resolved by the third author. To identify the location of noises, the research team added code that tracked the file names along with the line numbers that contained those noises.

Figure \ref{fig:longconditional} presented a long conditional code smell obtained from the \textit{user.py} file in the horizon project of \textit{OpenStack}. Our tool successfully flagged it as along conditional smell. We detected similar smells for the \textit{OpenStack} project that would help the cloud engineers to refactor those smells before they applied \textit{OpenStack} in real life. We ran the smell detector tool 4 times and each time successfully detected the same set of smells.

\begin{figure*}[htbp]
\centering
\subfloat{\includegraphics[width=0.35\textwidth]{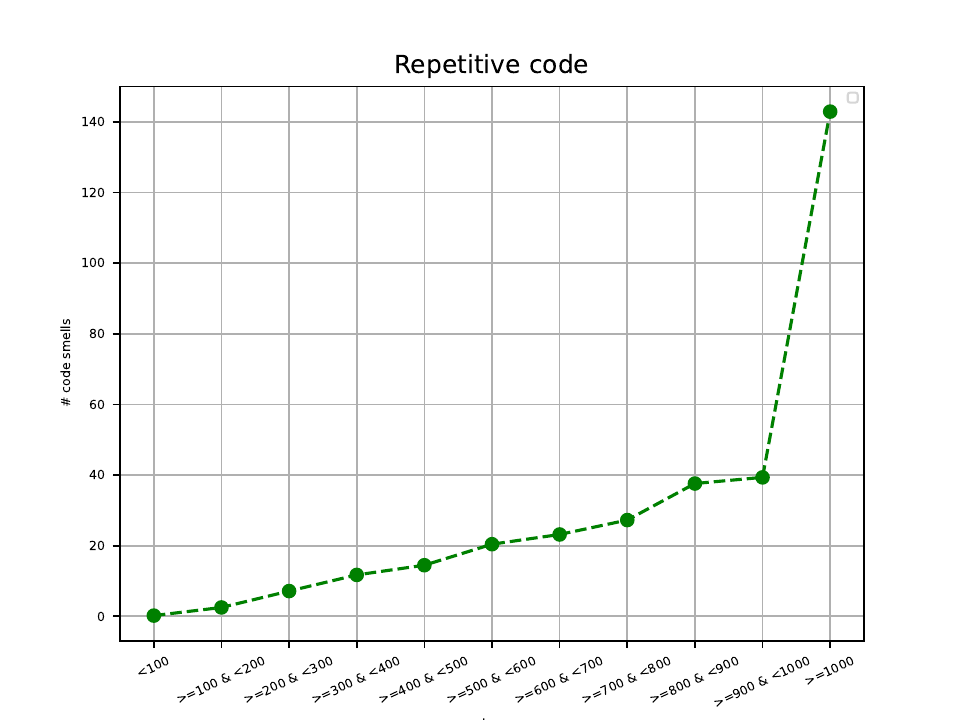}}
\subfloat{\includegraphics[width=0.35\textwidth]{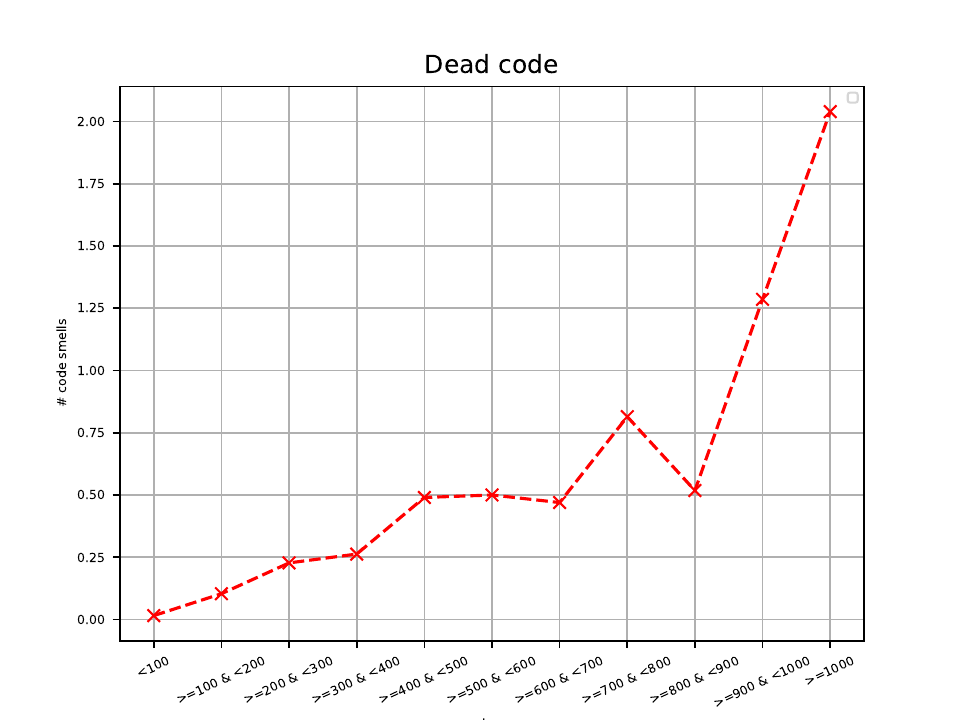}}
\subfloat{\includegraphics[width=0.35\textwidth]{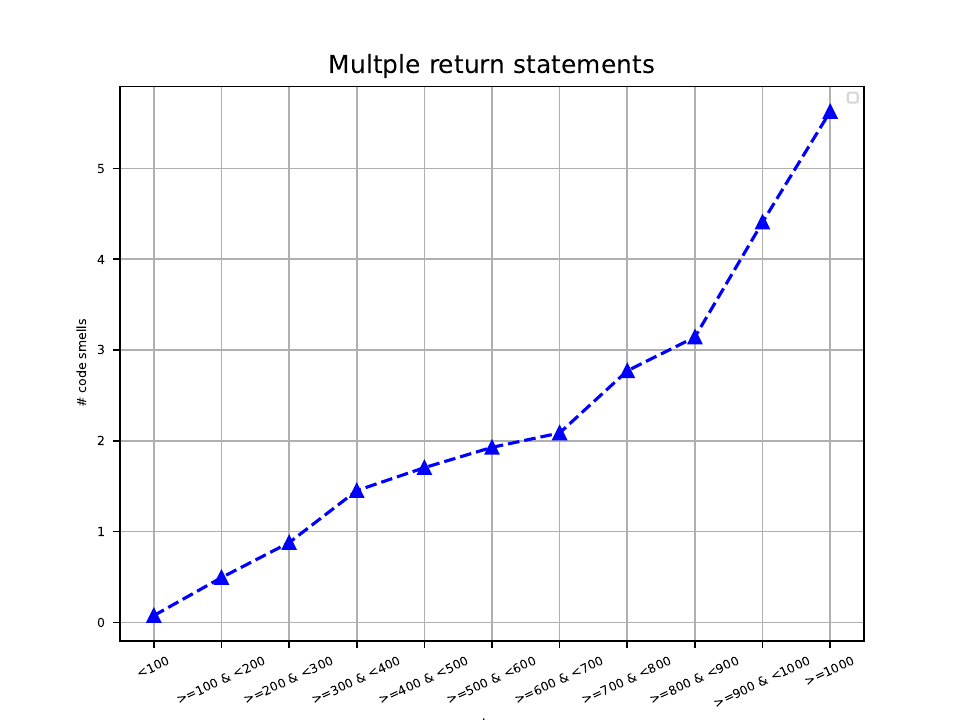}}

\subfloat{\includegraphics[width=0.35\textwidth]{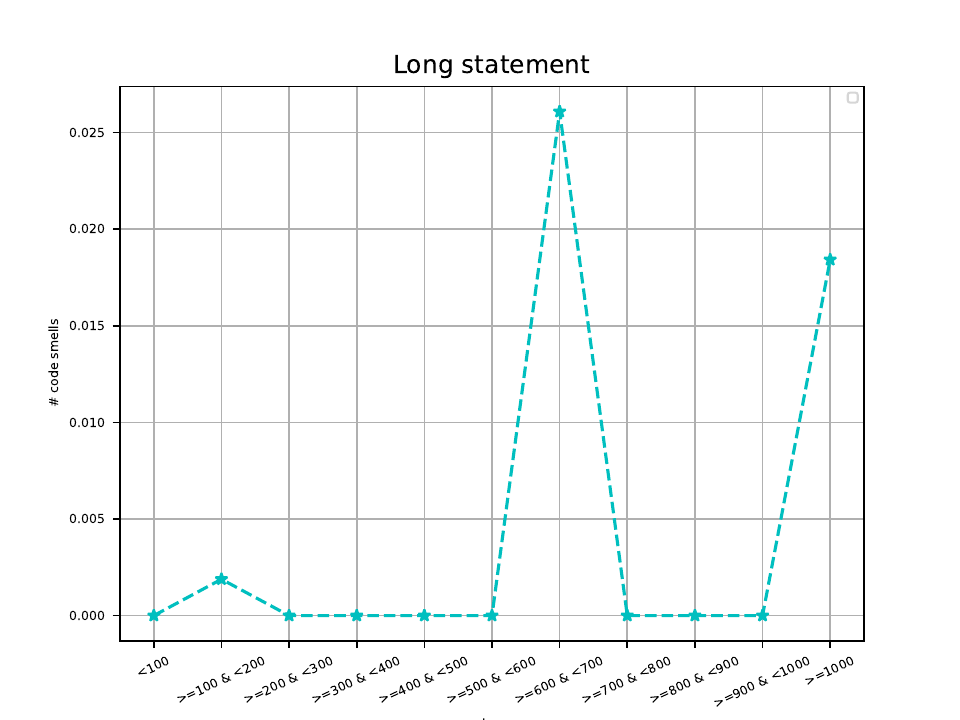}}
\subfloat{\includegraphics[width=0.35\textwidth]{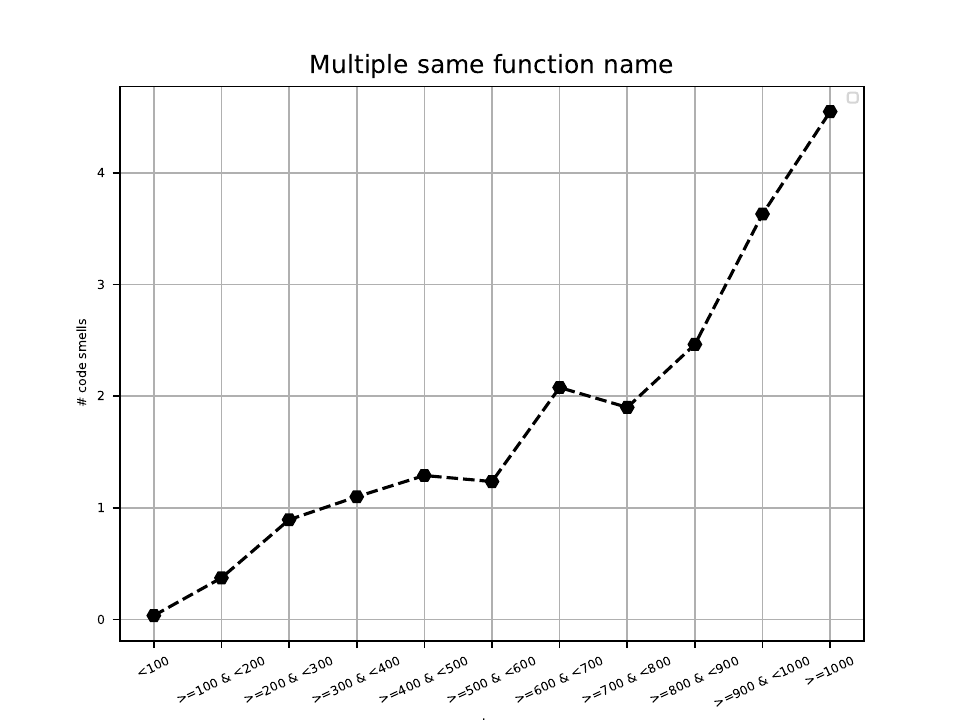}}
\subfloat{\includegraphics[width=0.35\textwidth]{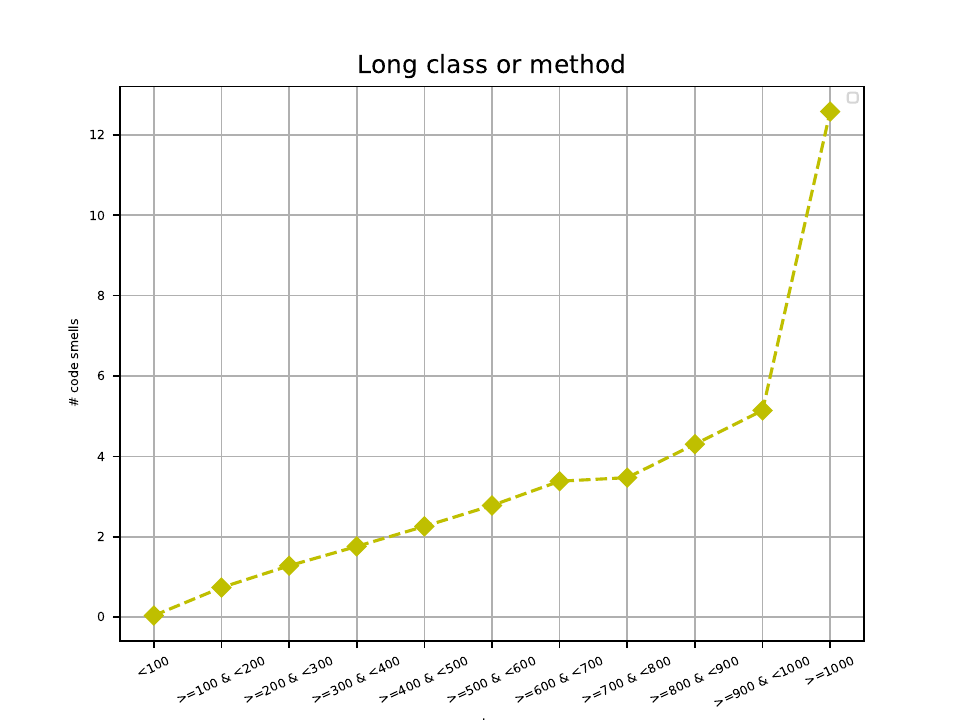}}

\subfloat{\includegraphics[width=0.35\textwidth]{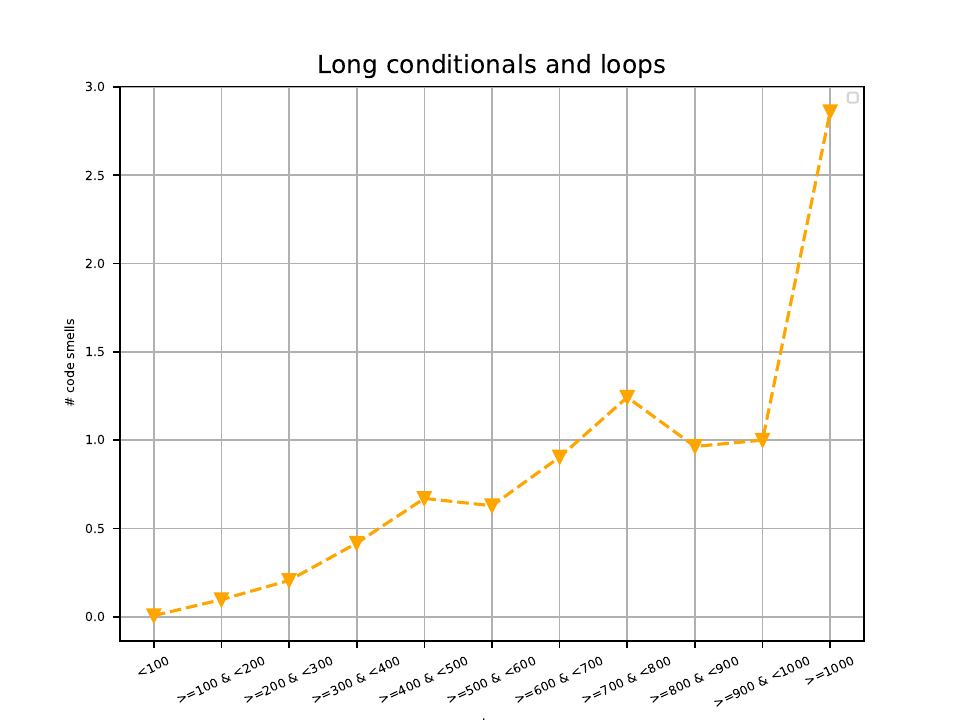}}
\subfloat{\includegraphics[width=0.35\textwidth]{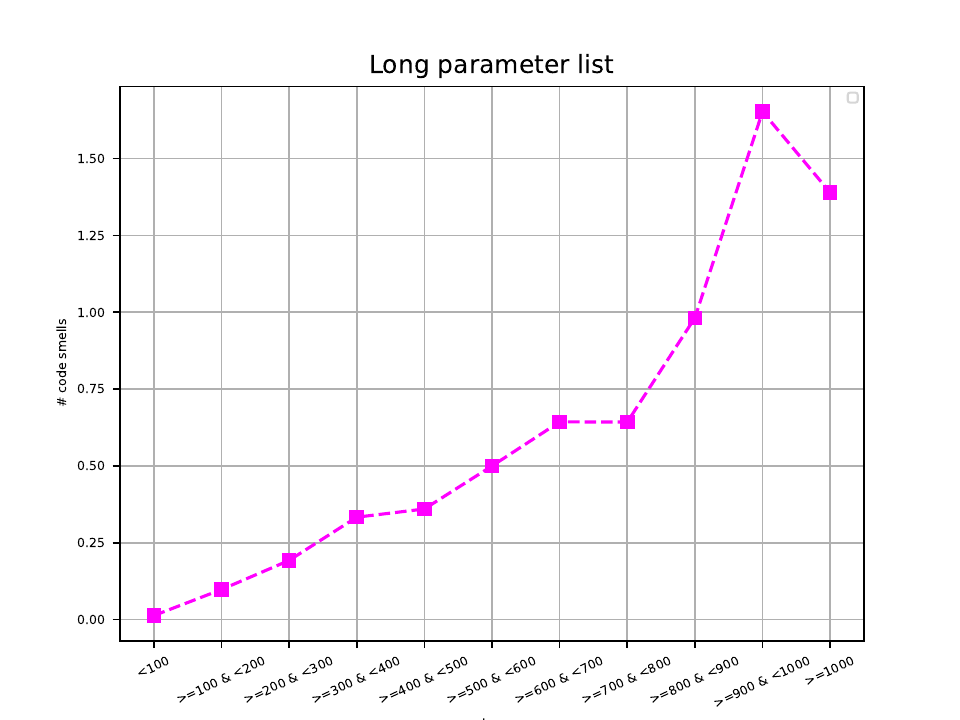}}
\caption{Obtained results of code smells detection in OpenStack cloud using CloudScent.}
\label{fig:result}
\end{figure*}

\begin{table}[!hbt]
{%
\begin{tabular}{|p{4.0cm}|p{1.8cm}|p{1.8cm}|}

\hline
 \textbf{lines of code (loc)}& \textbf{Total files=1788937)}& \textbf{normalized count}  \\
 \hline

Repetitive Codes & 78921 & 13.220\tabularnewline
\hline 
Dead Codes & 1492 & 0.254\tabularnewline
\hline 
Multiple Return Statements & 5195 & 0.873\tabularnewline
\hline 
Long Statements & 12 & 0.002\tabularnewline
\hline 
Multiple Same Function Names & 4109 & 0.690\tabularnewline
\hline 
Long Classes Or Methods & 8789 & 1.472\tabularnewline
\hline 
Long Conditionals or Loops & 1951 & 0.327\tabularnewline
\hline 
Long Parameter List & 1263 & 0.212\tabularnewline
\hline 
\end{tabular}}
\label{tab:summary}
\caption{Normalized data of detected smells using CloudScent.}
\end{table}

We wrote a python script that implemented \textit{CloudScent} to conduct the following tasks. We analyzed the code and added a row storing the sum of individual code smells. We created a table with the detected smells which showed the total number of smells. We created a column indicating the number of uncommented \textit{loc} in a file in range intervals of 100's ( 100, $\geq$100 \& $<$ 200, and so on) until 1000. We added one more column for files with \textit{loc} $>$1000. Each row indicated the sum of counts of different types of code smell in each column based on the size of the files. Then we added another column for getting a sum of counts of code smells in each row that helped to identify the individual smells. This helped us to verify that the total number of counts of code smells in the table was correct. 

Afterward, we added a new row that indicated the “number of files” for storing the count of files in each column based on the size of the files. We used the current table for all counts except for the last row and last column, we created another table for “Code smells average count per File” using the last row called the “number of lines”.  The data in this table was used to plot the number of smells which was normalized for the total size of the files.

The obtained results are shown in Figure \ref{fig:result}. We show the results for 8 code smells studied in this paper. The y-axis shows the number of code smells normalized by un-commented \textit{loc}. The x-axis represents the \textit{loc}. The experiments are repeated 4 times and in each case, we obtain the same results. We see that smells like repetitive code increases when the \textit{loc} for a file in \textit{OpenStack} increases above 900. Similar results are observed for \textit{dead code, multiple return statements, long classes, long conditionals, and long loops}. For \textit{multiple same function name} smell, the trend is seen even earlier when the \textit{loc }is 600. Table II focuses on the normalized results detected in the total number of files parsed by \textit{CloudScent}.

It is seen that the normalized number of code smells is proportional to the size of the python files in \textit{OpenStack}. Different behavior is noticed for long statement code smell, however, a closer look at the y-axis shows that the normalized value for this smell is negligible.

\section{Threats to validity}
\label{threats}
We identify the common threats to validity as experienced by similar qualitative research. Generalization of the survey results beyond the scope of open-source cloud platforms like \textit{OpenStack} needs to be conducted with care. We used rules to detect code smells in \textit{OpenStack}, and we need to generalize the results of the findings by testing on other cloud platforms like \textit{OpenNebula} \cite{milojivcic2011opennebula}. 

Our findings need to be evaluated by industry experts and also we should incorporate developer feedback to determine the accuracy of our results. Furthermore, we tested our model on only one version of \textit{OpenStack} which is the Zed series, we should also test it on different versions of \textit{OpenStack} in the future.

\section{Conclusion}
\label{conclusion}
This paper provided a model called \textit{CloudScent} to detect code smells in the open-source cloud. We implemented the tool in python and made it available as open-source. We encourage cloud software engineers to obtain our tool and detect smells in their cloud platform. We explored the different smells in open source cloud and provided syntactic tools to detect the smells. We showed the applicability of the approach by designing and developing an open-source tool that detected 8 different code smells in the open-source cloud, and we evaluated our findings with one industrial case study of \textit{OpenStack}.

To clarify the smells, our future research is focusing on extending the model to suggest refactoring for all the detected smells, with specific attention not to break the functionality as those refactoring are implemented. We also aim to identify the causes of those smells occurring in the cloud. We also aim to modify our model to be able to detect more code smells in the cloud. We look to test our tool on other case studies like the \textit{OpenNebula} project. At the same time, we plan to engage the cloud industry experts in our research to validate our proposed model. In the future, we plan to build a generalized model to predict code smells that may occur across various platforms of open-source clouds by using syntactic techniques and machine learning.


\bibliographystyle{ieeetr}
{\small
\bibliography{example}}

\begin{thebibliography}{10}

\bibitem{armbrust2010view}
M.~Armbrust, A.~Fox, R.~Griffith, A.~D. Joseph, R.~Katz, A.~Konwinski, G.~Lee,
  D.~Patterson, A.~Rabkin, I.~Stoica, {\em et~al.}, ``A view of cloud
  computing,'' {\em Communications of the ACM}, vol.~53, no.~4, pp.~50--58,
  2010.

\bibitem{antal2018hands}
G.~Antal, A.~Szarka, and P.~Heged{\H{u}}s, ``A hands-on openstack code
  refactoring experience report,'' in {\em Computational Science and Its
  Applications--ICCSA 2018: 18th International Conference, Melbourne, VIC,
  Australia, July 2-5, 2018, Proceedings, Part V 18}, pp.~464--480, Springer,
  2018.

\bibitem{10.1145/3571697.3571704}
A.~Imran and T.~Kosar, ``Qualitative analysis of the relationship between
  design smells and software engineering challenges,'' in {\em Proceedings of
  the 2022 European Symposium on Software Engineering}, ESSE '22, (New York,
  NY, USA), p.~48–55, Association for Computing Machinery, 2023.

\bibitem{tahir2020large}
A.~Tahir, J.~Dietrich, S.~Counsell, S.~Licorish, and A.~Yamashita, ``A large
  scale study on how developers discuss code smells and anti-pattern in stack
  exchange sites,'' {\em Information and Software Technology}, vol.~125,
  p.~106333, 2020.

\bibitem{han2021understanding}
X.~Han, A.~Tahir, P.~Liang, S.~Counsell, and Y.~Luo, ``Understanding code smell
  detection via code review: A study of the openstack community,'' in {\em 2021
  IEEE/ACM 29th International Conference on Program Comprehension (ICPC)},
  pp.~323--334, IEEE, 2021.

\bibitem{tsantalis2018ten}
N.~Tsantalis, T.~Chaikalis, and A.~Chatzigeorgiou, ``Ten years of jdeodorant:
  Lessons learned from the hunt for smells,'' in {\em 2018 IEEE 25th
  international conference on software analysis, evolution and reengineering
  (SANER)}, pp.~4--14, IEEE, 2018.

\bibitem{pmdsourcecodeanalyzer2022}
https://pmd.github.io/latest/index.html, ``Pmd,'' Oct 2022.

\bibitem{marcilio2019static}
D.~Marcilio, R.~Bonif{\'a}cio, E.~Monteiro, E.~Canedo, W.~Luz, and G.~Pinto,
  ``Are static analysis violations really fixed? a closer look at realistic
  usage of sonarqube,'' in {\em 2019 IEEE/ACM 27th International Conference on
  Program Comprehension (ICPC)}, pp.~209--219, IEEE, 2019.

\bibitem{moha2009decor}
N.~Moha, Y.-G. Gueheneuc, L.~Duchien, and A.-F. Le~Meur, ``Decor: A method for
  the specification and detection of code and design smells,'' {\em IEEE
  Transactions on Software Engineering}, vol.~36, no.~1, pp.~20--36, 2009.

\bibitem{kumara2020towards}
I.~Kumara, Z.~Vasileiou, G.~Meditskos, D.~A. Tamburri, W.-J. Van Den~Heuvel,
  A.~Karakostas, S.~Vrochidis, and I.~Kompatsiaris, ``Towards semantic
  detection of smells in cloud infrastructure code,'' in {\em Proceedings of
  the 10th International Conference on Web Intelligence, Mining and Semantics},
  pp.~63--67, 2020.

\bibitem{9118465}
N.~Almashfi and L.~Lu, ``Code smell detection tool for java script programs,''
  in {\em 2020 5th International Conference on Computer and Communication
  Systems (ICCCS)}, pp.~172--176, 2020.

\bibitem{muse2020prevalence}
B.~A. Muse, M.~M. Rahman, C.~Nagy, A.~Cleve, F.~Khomh, and G.~Antoniol, ``On
  the prevalence, impact, and evolution of sql code smells in data-intensive
  systems,'' in {\em Proceedings of the 17th international conference on mining
  software repositories}, pp.~327--338, 2020.

\bibitem{bessghaier2020diffusion}
N.~Bessghaier, A.~Ouni, and M.~W. Mkaouer, ``On the diffusion and impact of
  code smells in web applications,'' in {\em Services Computing--SCC 2020: 17th
  International Conference, Held as Part of the Services Conference Federation,
  SCF 2020, Honolulu, HI, USA, September 18--20, 2020, Proceedings 17},
  pp.~67--84, Springer, 2020.

\bibitem{rosado2014overview}
T.~Rosado and J.~Bernardino, ``An overview of openstack architecture,'' in {\em
  Proceedings of the 18th International Database Engineering \& Applications
  Symposium}, pp.~366--367, 2014.

\bibitem{wen2012comparison}
X.~Wen, G.~Gu, Q.~Li, Y.~Gao, and X.~Zhang, ``Comparison of open-source cloud
  management platforms: Openstack and opennebula,'' in {\em 2012 9th
  International Conference on Fuzzy Systems and Knowledge Discovery},
  pp.~2457--2461, IEEE, 2012.

\bibitem{apostu2013study}
A.~Apostu, F.~Puican, G.~Ularu, G.~Suciu, G.~Todoran, {\em et~al.}, ``Study on
  advantages and disadvantages of cloud computing--the advantages of telemetry
  applications in the cloud,'' {\em Recent advances in applied computer science
  and digital services}, vol.~2103, 2013.

\bibitem{milojivcic2011opennebula}
D.~Milojicic, I.~M. Llorente, and R.~S. Montero, ``Opennebula: A cloud
  management tool,'' {\em IEEE Internet Computing}, vol.~15, no.~2, pp.~11--14,
  2011.

\end{thebibliography}

\end{document}